\documentclass[prl,twocolumn,superscriptaddress,floatfix,longbibliography]{revtex4}%nofootinbib,floatfix

\usepackage{amsfonts}
\usepackage{amsmath}
\usepackage{graphicx}
\usepackage{amssymb}
\usepackage{color}
\usepackage{subfigure}
\usepackage{mathtools}
\usepackage{tikz-cd}

\newcommand{\avg}[1]{\big\langle#1\big\rangle}

\begin{document}
\title{Numerical studies of back-reaction effects in an analog model of cosmological pre-heating}

\author{Salvatore Butera}
\affiliation{School of Physics and Astronomy, University of Glasgow, Glasgow G12 8QQ, UK}
\author{Iacopo Carusotto}
\affiliation{INO-CNR BEC Center and Dipartimento di Fisica, Universit\`a di Trento, I-38123 Povo, Italy}
\begin{abstract}
We theoretically propose an atomic Bose-Einstein condensate as an analog model of back-reaction effects during the pre-heating stage of early Universe.
In particular, we address the out-of-equilibrium dynamics where the initially excited inflaton field decays by parametrically exciting the matter fields. We consider a two-dimensional, ring-shaped BEC under a tight transverse confinement whose transverse breathing mode and the Goldstone and dipole excitation branches simulate the inflaton and quantum matter fields, respectively. A strong excitation of the breathing mode leads to an exponentially-growing emission of dipole and Goldstone excitations via parametric pair creation: our numerical simulations of the BEC dynamics show how the associated back-reaction effect  not only results in an effective friction  of the breathing mode but also in a quick loss of longitudinal spatial coherence of the initially in-phase excitations. Implications of this result on the validity of the usual semiclassical description of back-reaction are finally discussed.

\end{abstract}
\maketitle

\emph{Introduction -- }
Since Unruh's pioneering proposal~\cite{Unruh-Analog-1981}, analog models of gravity represent a promising platform where a wide range of effects of quantum fields in curved spacetime can be studied from first principles and potentially find experimental confirmation~\cite{Barcelo-2011}.
A most celebrated achievement was the observation of the analog of Hawking radiation~\cite{Hawking1974,Hawking1975} emanating from the acoustic horizon in a trans-sonically flowing Bose-Einstein condensates (BECs) of ultra-cold atoms~\cite{Garay2000,Carusotto2008,steinhauer2016,steinhauer2019,steinhauer2021}. Analogs of cosmological particle creation effects have also been investigated on the BEC platform both theoretically~\cite{Fedichev2004,Fischer2004,Jain-Cosm-BEC-2007,Carusotto2010,Butera2021} and experimentally~\cite{Hung1213,Eckel-Cosm-BEC-2018}.

While these advances clearly demonstrate the power of the analog gravity program, they all address kinematic, test-field effects of a non-interacting quantum field theory on a pre-determined curved space-time background~\cite{birrell1984quantum}. The next challenge that stands in front of the analog gravity community is to extend these investigations to the so-called {\em back-reaction} phenomena~\cite{hu_verdaguer_2020}, where the background has its own dynamics and interplays with the quantum field. 

A simplest example of such an effect is the radiative friction felt by a accelerated mirror in response to the dynamical Casimir emission (DCE)~\cite{KardarRMP1999}, where theoretical studies in a single-mode geometry~\cite{Savasta-PRX-2018,Butera-BR_DCE-2019-PRA} have hinted at an important role of quantum fluctuations of the friction force~\cite{MaiaNeto-PRL-Decoher,MaiaNeto-PRA-Decoher,Butera-BR_DCE-2019-EPL}. Beyond these toy-models, a full understanding of quantum features in back-reaction effects is of outstanding importance in the case of black hole evaporation under the effect of Hawking emission~\cite{Fabbri-book}, where one expects that quantum fluctuations may be involved in the so-called information paradox~\cite{Maldacena-InfoParadox-2020}. First pioneering steps in this direction have been moved in~\cite{DeNova:PRA16,Robertson-PRD-PreHeatAn-2019,Patrick:PRL21}.

A seemingly insurmountable hurdle in extending analog gravity towards back-reaction effects is posed by the starkly different form of the nonlinear evolution equations, governed by Einstein gravity and the (still unknown) physics of space-time at the Planck scale on one side and by the (well controlled) microscopic material dynamics of the analog model on the other side. In this work, we fully acknowledge this difficulty and, as a workaround, we adopt a phenomenological perspective based on the working assumption that qualitative, mesoscopic observable effects of back-reaction such as dissipation, fluctuation and decoherence are universal, since they result from a coarse-graining process, and are thus expected to be ultimately insensitive to the microscopic details of the interactions~\cite{BeiLok-CondSpacetime}. A similar perspective also underlies the widespread use of phenomenological models of the inflaton and of its coupling to matter and gauge fields in the cosmological literature.

Moving along these lines, we consider in this Letter an analog model of the pre-heating of the early Universe \cite{Inflation-rev,Brandenberger-rev,Amin-rev}. This is the later stage of the inflation, when the inflaton field has ended its slow-rolling on the potential plateau and has fallen into the final potential well. The ensuing periodic oscillations around the bottom of the potential well parametrically excite the vacuum fluctuations of the matter fields that are coupled to the inflaton, resulting in an explosive production of matter in the Universe and a corresponding decay of the inflaton oscillations \cite{Kofman-prd-1994,Brandenberger-prd-1995,Tkachev-prl-1996,Son-prl-1996,Allahverdi-prd-1997}.

In our analog model, we simulate this dynamics by using a two-dimensional ring-shaped BEC configuration first proposed in this context in~\cite{Robertson-PRD-PreHeatAn-2019}: the oscillations of the inflaton field are implemented by exciting a breathing mode along the transverse direction. These relatively high energy oscillations then lead to the parametric amplification of zero-point fluctuations in the lower energy longitudinal (dipole and Goldstone) modes, which provide an analog for the matter fields. Our study is based on {\em ab initio} numerical simulations of the atomic cloud dynamics via the so-called truncated Wigner approximation (TWA)~\cite{Steel-PRA-Wigner-1998,Sinatra-PRA-Wigner-2001}: in contrast to the semiclassical treatment of the breathing mode dynamics in~\cite{Robertson-PRD-PreHeatAn-2019}, our approach is able to include the dynamical interplay of the quantum fluctuations in the different modes via the back-reaction effects. Differently from~\cite{Hauke-preheating-2021}, we will not dwelve here into the physics of thermalization of the generated particles during the successive re-heating stage~\cite{Kofman:PRD97,Roos-PRD-1997,Micha:PRL03,Figueroa_2017}.

\vspace{2mm}
\emph{The system. }
We consider a dilute two-dimensional gas of mass $m$ atoms at zero temperature, homogeneous along the longitudinal $x$ direction with periodic boundary conditions and trapped in the transverse $y$ direction by an external potential $V_{\rm ext}(y)$. For numerical convenience, this is taken as harmonic of frequency $\omega_0$ at small $y$ with a hard-wall at $y=\pm L_y/2$ on both sides.
%Because of computational reasons, the numerical box that we use to run our simulations is finite, and the system is ultimately limited in the transverse direction by a hard-wall potential. 
The many-body Hamiltonian reads \cite{BEC-Rev}:
\begin{equation}
	\hat{H} = \int{d\mathbf{r} \left[\hat{\Psi}^\dag(\mathbf{r}) \hat{h} \hat{\Psi}(\mathbf{r}) + \frac{U}{2}\hat{\Psi}^\dag(\mathbf{r})\hat{\Psi}^\dag(\mathbf{r})\hat{\Psi}(\mathbf{r})\hat{\Psi}(\mathbf{r})\right]},
\label{Eq:H}
\end{equation}
where $\hat{h} \equiv -(\hbar^2/2m)\boldsymbol{\nabla}^2 + V_{\rm ext}(y)$ is the single particle Hamiltonian and $U$ is the strength of the zero-range inter-atomic collisional interaction. At the mean-field level, the evolution of the gas is captured by the standard Gross-Pitaevskii equation (GPE) for the order parameter~\cite{Stringari_BEC}.

\begin{figure}[t!]
\centering
\includegraphics[width=0.45\textwidth]{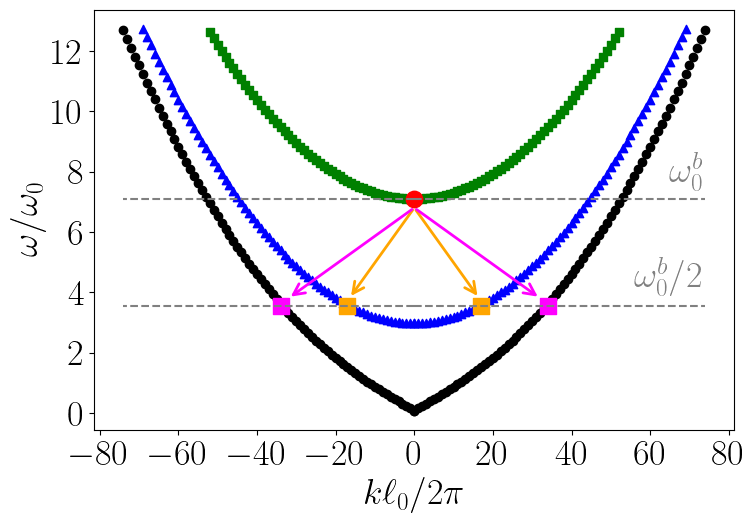}
\caption{Bogoliubov spectrum of collective excitations around the ground state. The three curves correspond to modes with zero (Goldstone, black), one (dipole, blue) and two (breathing, green) nodes in the transverse direction. The red circle highlights the transverse breathing mode that is excited at early times to simulate the inflaton oscillations; the yellow (purple) circles highlight the dipole (Goldstone) modes of opposite momenta that are resonantly excited by the parametric processes indicated by the arrows. System parameters: gas of $N=10^6$ atoms in an integration box of size $L_{x,y}/\ell_0 = 140, 3.54$ in units of the transverse harmonic oscillator length $\ell_0=\sqrt{2\hbar/m\omega_0}$, with $N_{x,y}=512,12$ grid points; equilibrium chemical potential $\mu/\hbar\omega_0= 2.38$.}
\label{Fig:1}
\end{figure}

Quantum fluctuations and the dynamics of small  excitations  around the mean-field state can be described at second order in the fluctuation amplitude by the Bogoliubov theory~\cite{Castin-BogNumCons-1998}. The spectrum $\{\omega_n^r\}$ %$(n\in\mathbb{N})$ 
of the collective Bogoliubov modes on top of a stationary ground-state BEC and the corresponding eigenfunctions $\left\{u_n^r,v_n^r\right\}$ are calculated by diagonalizing the Bogoliubov operator $\mathcal{L}_{\rm Bog}$~\cite{Castin-BogNumCons-1998} within a suitable integration box of size $L_{x,y}$ including $N_{x,y}$ grid-points in each direction. For each mode, the integer-valued subscript $n$ and the superscript $r=g,\,d,\,b,\,...$ respectively identify the longitudinal wave vector $k=2\pi n/L_x$  and the different excitation branches, labelled by the number of transverse nodes in the wavefunction ($g=\text{Goldstone}$ -- 0 nodes; $d=\text{dipole}$ -- 1 node; $b=\text{breathing}$ -- 2 nodes). As a concrete example, the dispersion of the Goldstone, dipole and breathing branches for the system parameters used throughout this work is shown in Fig.~\ref{Fig:1}, together with the basic parametric emission processes. Because of the anharmonicity of the transverse confinement potential, parametric emission can occur in both $g$ and $d$ branches. 

The usual way of understanding the parametric emission of Bogoliubov quanta in time-dependent condensates consists of generalizing the Bogoliubov theory to the case of a time-dependent Bogoliubov operator $\mathcal{L}_{\rm Bog}(t)$~\cite{Carusotto-PRA-AnBackR-2012,cominotti2022observation}. While this approach well describes both spontaneous and stimulated (and thus exponentially growing) parametric emission processes, it implicitly assumes that the background dynamics encoded in $\mathcal{L}_{\rm Bog}(t)$ is not affected by the parametric emission. In order to capture the back-reaction effect of the parametric emission onto the breathing mode oscillations and, in particular, its quantum fluctuations, we need to go beyond this mean-field-like picture.

\emph{Simulation method.} In our alternative picture, the parametric emission can be seen as the conversion of $b$-branch Bogoliubov quanta of the ground state condensate into pairs of quanta in either the $g$ or the $d$ branches, mediated by nonlinear cubic terms that go beyond the quadratic Bogoliubov Hamiltonian and describe interactions and inter-conversion between the Bogoliubov modes, e.g. the so-called Beliaev-Landau damping processes~\cite{pitaevskii1997landau,Liu:PRL1997,Giorgini:PRA1998,Fedichev:PRA98}.

\begin{figure*}[!t]
    \centering
		{\includegraphics[width=0.32\textwidth]{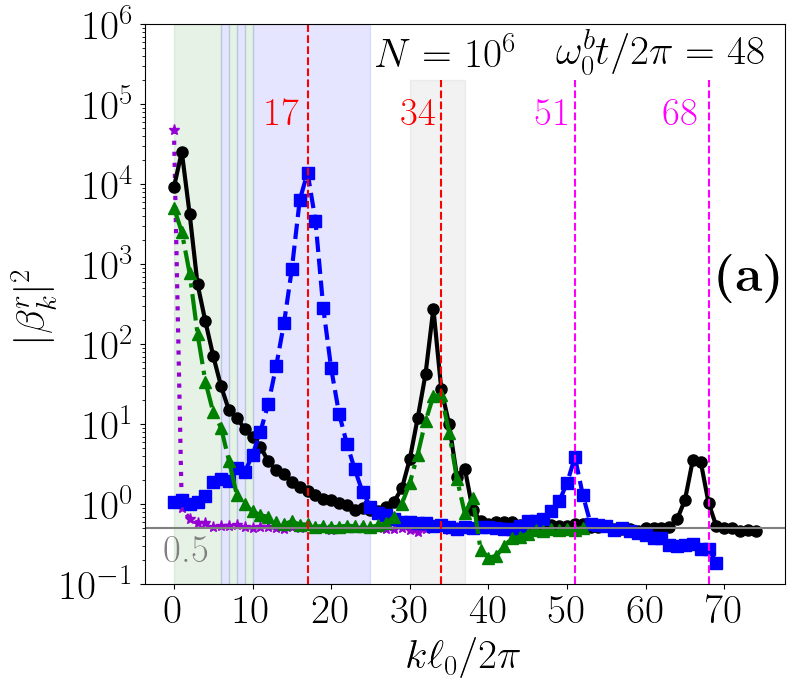}}
		{\includegraphics[width=0.32\textwidth]{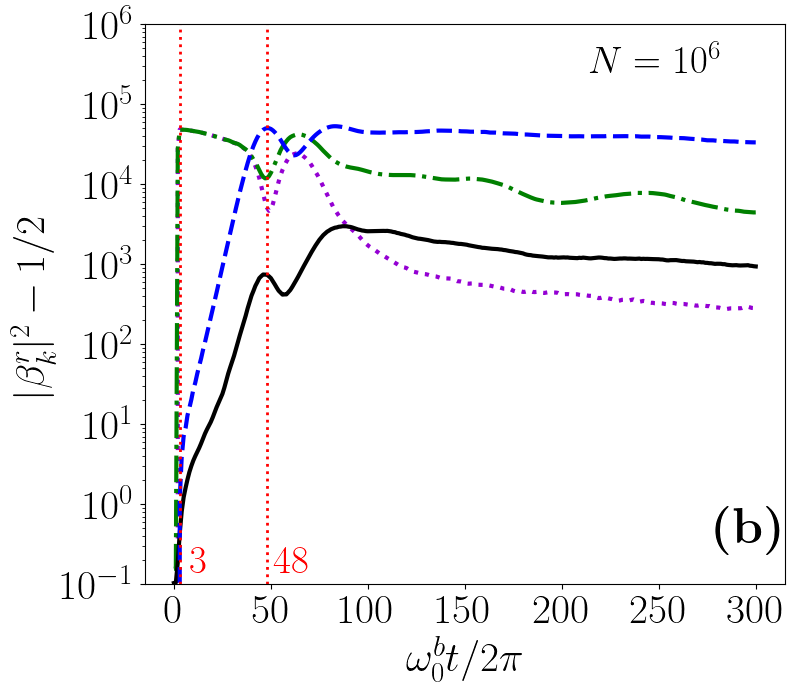}}
		{\includegraphics[width=0.32\textwidth]{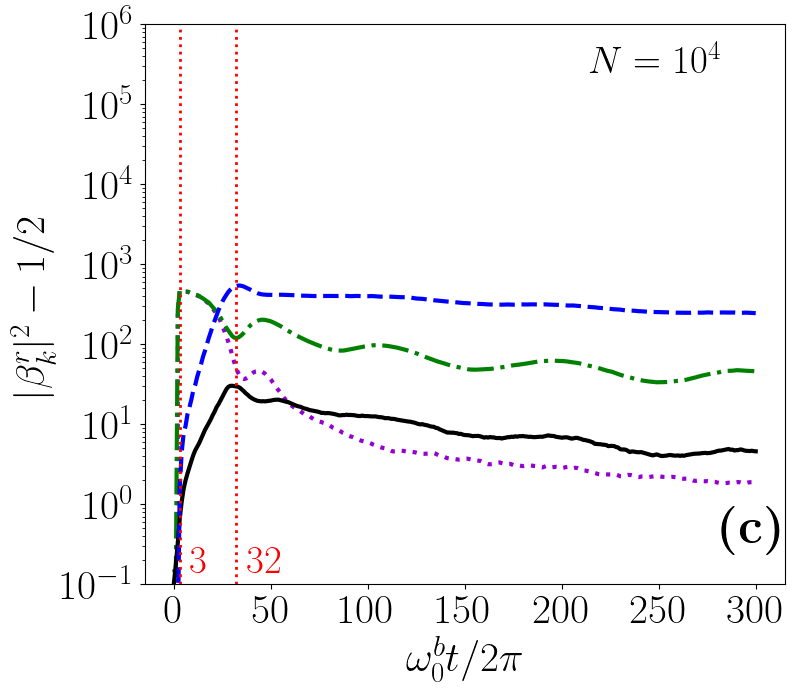}}		
		\caption{Numerical results of the TWA simulations. Stochastic averages are based on a sample of $\mathcal{N}_r=1000$ independent realization. Panel (a): Momentum distribution of the population in the Bogoliubov modes at time $\omega_0^b t/2\pi = 48$. The dot-dashed green, dashed blue and solid black lines correspond to the breathing, dipole and Goldstone branches. The dotted violet line shows the population in the breathing mode immediatly after the excitation sequence. Panels (b,c): time-evolution of the integrated population in the breathing, dipole, and Goldstone branches over the regions indicated by the shading in panel (a). Same color code as in (a). The dotted violet line shows the time-evolution of the population in the single breathing mode at $k=0$. Panel (c): same calculation as in (b) for a reduced atom number $N=10^4$ at a fixed $\mu$. }
\label{Fig:2}
\end{figure*}

Beyond this perturbative picture, a simulation of the full nonlinear dynamics of the system including non-perturbative interactions between Bogoliubov quasi-particles can be numerically carried out within the TWA~\cite{Steel-PRA-Wigner-1998,Sinatra-PRA-Wigner-2001}. Such an approach has proven suitable for studying quantum field effects at the test field level such as analogs of cosmological particle creation~\cite{Jain-Cosm-BEC-2007} and of Hawking radiation~\cite{Carusotto2008}, and has recently started to be pushed beyond this regime~\cite{Robertson-PRD-PreHeatAn-2019}. Interestingly, similar classical field approaches are of current use also in the cosmological literature~\cite{Roos-PRD-1997}. The basic idea of TWA is to describe the quantum field operator $\hat{\Psi}(\mathbf{r})$ in terms of a suitably distributed stochastic classical field  $\psi(\mathbf{r})$. 
Quantum expectation values of symmetrically-ordered observables are then obtained as stochastic averages of the corresponding classical field objects,
$\avg{\mathcal{S}[(\hat{\Psi}^\dag)^q\,\hat{\Psi}^p\,]} = \avg{(\psi^*)^q\,\psi^p }_{W}$. 

For conservative systems like atomic gases, the classical field $\psi(\mathbf{r})$ follows a deterministic time evolution according to a standard GPE and quantum noise is encoded in the initial state of the system. In particular, we consider the gas to be initially in its ground state; under the assumption of a dilute gas, this is accurately described as the ground state of the Bogoliubov theory. Within the TWA, the classical field corresponding to such a state is constructed as the sum of the GPE ground state $\phi(\mathbf{r})$, plus a Gaussian stochastic component that accounts for the zero-point fluctuations of the Bogoliubov modes~\cite{Steel-PRA-Wigner-1998,Sinatra-PRA-Wigner-2001},
\begin{equation}
	\psi\left(\mathbf{r},t=0\right)=\phi(\mathbf{r})+\sum_{n,r}{\left(\beta_n^r u_n^r(\mathbf{r})+{\beta_n^r}^*{v_n^r}^*(\mathbf{r})\right)},
\label{Eq:InitialState}
\end{equation}
where the $\beta_n^r$ coefficients for each (positive-norm) $n,r$ Bogoliubov mode are independent, zero-mean, Gaussian random variables with $\big<{\beta_k^2}\big>=0$ and $\big<{|\beta_k|^2}\big>=1/2$. In our simulations, a sample of $\mathcal{N}_r=1000$ independent realizations of the classical field is used, with an initial distribution drawn according to Eq.~\eqref{Eq:InitialState}. 

\begin{figure*}[!t]
\centering
{\includegraphics[width=0.35\textwidth]{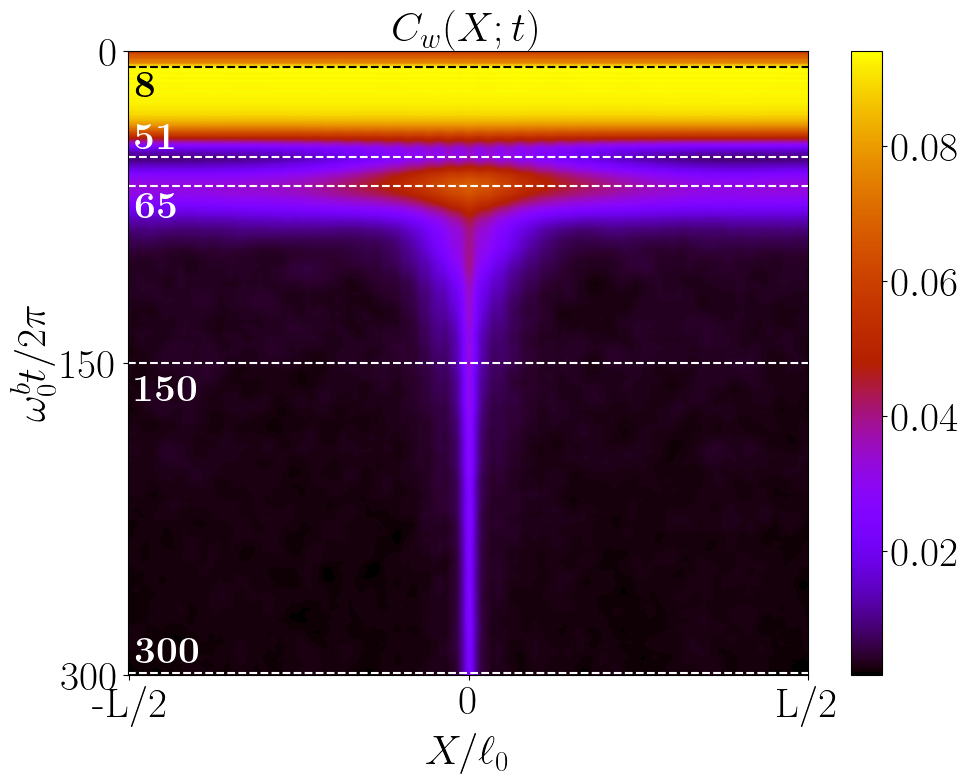}}
{\includegraphics[width=0.31\textwidth]{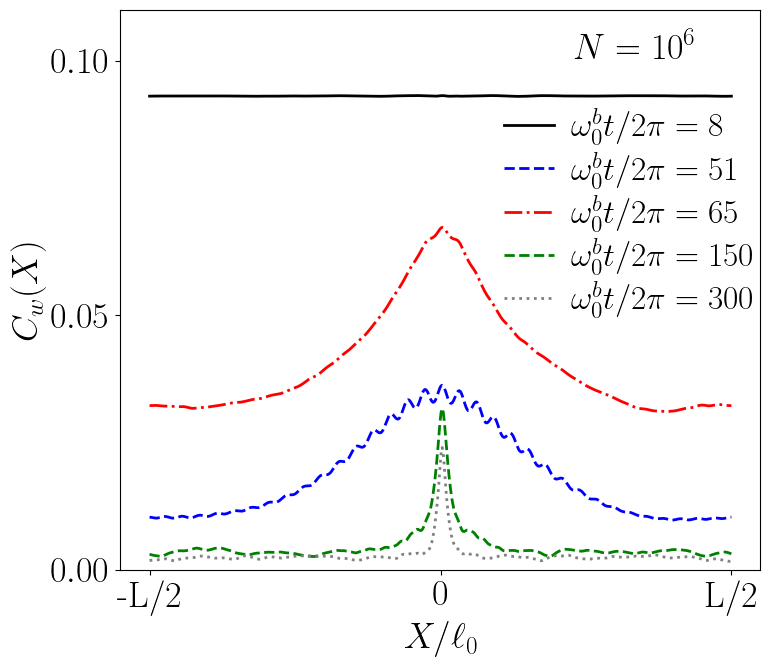}}
{\includegraphics[width=0.31\textwidth]{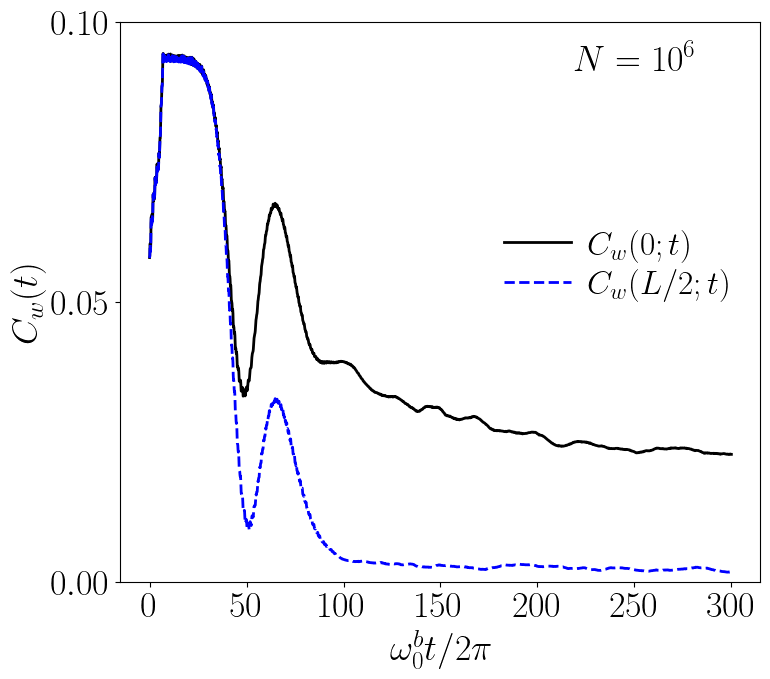}}
\caption{Left panel: Time-evolution of the spatial correlation function $C_{w}(X;t)$ of the transverse size of the condensate. Cuts at different times (indicated by the horizontal dashed lines) are shown in the middle panel. Right panel: time evolution of the spatial correlation function for coincident $C_{w}(0;t)$ and opposite $C_{w}(L/2;t)$ points around the ring: the former gives information on the total oscillation intensity, while the latter indicates the spatial coherence of the oscillations. Same system and simulation parameters as in Fig.\ref{Fig:1} and \ref{Fig:2}.}
\label{Fig:3}
\end{figure*}

\emph{Numerical results.} 
In Fig.~\ref{Fig:2}, we illustrate the evolution of the population in the different Bogoliubov modes, $n_n^r(t)+1/2\equiv \big<{(\hat{b}^r_n)^\dag\,\hat{b}^r_n\, + \hat{b}^r_n\, (\hat{b}^r_n)^\dag\,}\big>/2 = \avg{|\beta^r_n(t)|^2}_W$ as a function of time. 
At the initial time $t=0$, the system is prepared in the Bogoliubov vacuum and the modes only host zero-point fluctuations, $n_n^r(t=0)=0$. Around a time $\omega_0 t_0 /2\pi = 2$, we impart a short modulation of the trapping frequency, 
%\begin{equation}
	$\omega_0(t)/\omega_0 = 1 + A e^{-(t-t_0)^2/2\sigma_t^2}$
%\label{Eq:FreqMod}
%\end{equation}
of amplitude $A$ and duration $\sigma_t$ so to excite the spatially-uniform $k=0$ transverse breathing mode of the condensate (red circle in Fig.~\ref{Fig:1}). This excitation is visible as a marked $k=0$ peak in the momentum-resolved occupation of the $b$ modes right after the kick [purple dotted line in Fig.~\ref{Fig:2}(a)].

Afterwards, the nonlinear coupling between the $b$ and $g,d$ modes makes the vacuum fluctuations in the dipole and Goldstone modes to get parametrically excited by the oscillations in the transverse direction. Because of the ring configuration here considered, momentum in the longitudinal direction is conserved and parametric down-conversion process involves pairs of particles with opposite momenta as  indicated by arrows in Fig.~\ref{Fig:1}. Energy conservation makes the parametric processes to be most effective into the Goldstone and dipole modes of frequency $\omega_{\rm res}^{g,d} =\omega_0^b/2$ for which the parametric emission is resonant with the breathing mode oscillations at $\omega_0^b$ that are driving it, as shown in the momentum distributions of Fig.~\ref{Fig:2}(a). 

Given the bosonic nature of the Bogoliubov modes, the parametric emission starts from zero-point quantum fluctuations but then gets self-stimulated as more and more population is created, leading to an exponential growth of the population in the resonant modes. This behaviour is apparent in the black-solid and blue-dashed lines of Fig.~\ref{Fig:2}(b), which display the evolution of the integrated population within the integration windows indicated by the shaded areas in Fig.~\ref{Fig:2}(a). This exponential growth is the analogous of the explosive production of matter that takes place during the pre-heating of the early Universe. 

Once the population in the $g,d$ modes has grown to sizable values, nonlinear and saturation effects start taking place~\cite{Kofman:PRD97,Roos-PRD-1997,Micha:PRL03,Figueroa_2017}. Nonlinear effects are visible in the appearance of harmonic peaks in the momentum distributions shown in Fig.~\ref{Fig:2}(a)~\cite{Micha:PRL03,Robertson-PRD-PreHeatAn-2019,Silke_new} as well as an increased width of all peaks. Self-interaction, scattering and thermalization processes within the Goldstone and dipole branches are responsible for these effects: while they are of great interest as an analog model~\cite{Hauke-preheating-2021} of cosmological reheating~\cite{Kofman:PRD97,Roos-PRD-1997,Micha:PRL03,Figueroa_2017}, a detailed study goes beyond the purpose of this work and is postponed to future work.

Here, we rather focus on the dynamics of the breathing mode $b$ during the pre-heating stage. In Fig.~\ref{Fig:2}(b), we see a marked drop in the integrated population in the breathing $b$ branch (green-dashed line) as well as a saturation of the parametric emission rate as soon as the populations in the $g,d$ modes have grown to a value comparable to the $b$ mode and back-reaction effects have started exerting a sizable effective friction onto the $b$ mode. Interestingly, such a damping is not purely monotonic, but energy gets at least partially exchanged between the transverse and longitudinal modes. This intermediate-time damped-oscillatory phenomenology is qualitatitively similar to the one predicted in~\cite{Butera-BR_DCE-2019-PRA,Butera-BR_DCE-2019-EPL} for the back-reaction effect of dynamical Casimir processes in a single-mode cavity configuration, with interesting new features stemming from the many-mode nature of our system. The quantum fluctuation origin of the friction effect is confirmed by the plot in Fig.~\ref{Fig:2}(c): quantum fluctuations are more important in stronger interacting systems with a larger interaction constant $U$ at fixed chemical potential $\mu$, which results in a stronger friction and less marked oscillations.

\emph{Local observables. } One may be concerned that the very concept of Bogoliubov modes may cease being well defined in a regime where nonlinear effects start playing a major role. To circumvent this objection and, at the same time, open a new perspective on this physics, we complement the mode-wise analysis of Fig.~\ref{Fig:2} with a study of real-space quantities that enjoy a direct definition in terms of the real-space classical field $\psi(\mathbf{r},t)$. Specifically, we consider the $x$-dependent transverse cloud size as
\begin{equation}
    w(x,t) \equiv \frac{\int_0^{L_y} dy |\psi(\mathbf{r},t)|^2 y^2}{\int_0^{L_y} dy |\psi(\mathbf{r},t)|^2},
    \label{Eq:w}
\end{equation}
and the spatial correlation function 
\begin{equation}
	C_{w}(X;t) \equiv \Big< \frac{\delta w(x,t)\delta w(x+X,t)}{\bar{w}^2(0)}\Big>_W,\label{Eq:VarianceCorr}
\end{equation}
of its fluctuations $\delta w(x,t) = w(x,t) - \bar{w}(0)$ from the initial value $\bar{w}(0)$ of its spatial average $\bar{w}(t) \equiv L_x^{-1}\int_0^{L_x}dx\, w(x,t)$ at $t=0$ . The time evolution of $C_{w}(X;t)$ is illustrated in the left panel of Fig.~\ref{Fig:3}: Right after the initial kick around $\omega_0^b t=2$, the breathing mode oscillates coherently with a uniform amplitude throughout the cloud, and the correlation function is large and constant. Then, not only the overall magnitude of $C_{w}$ decreases as a signature of back-reaction-induced damping, but spatial coherence is also lost at an even faster rate. 

As one can see in the cuts in the middle panel, at late times the correlation function maintains a sizable value only in the sharp peak around $x=0$: referring specifically to the $\omega_0^b t= 150$ curve, this indicates that the intensity of the $b$ oscillations, encoded in $C_{w}(0;t)$, has dropped by a factor around $3$, but has almost completely lost its spatial decoherence as shown by the negligible value $C_{w}(L/2;t)\approx 0$. A further visualization of this effect is available in the right panel of Fig.~\ref{Fig:3}, which highlights the much faster decrease of the long-distance coherence $C_{w}(L/2;t)$ (blue dashed line) and a relative stabilization of the oscillation intensity (black solid line). A momentum-space signature of this effect is visible as a marked broadening of the $k$-space breathing mode distribution as time proceeds [green line in Fig.~\ref{Fig:2}(a)], whose large momentum-width corresponds to the short real-space coherence length. This population redistribution from the single initial $k=0$ mode into a wider band of modes is further visible by comparing the purple-dashed and green-dot-dashed curves in panels (b,c), whose behaviour in time recovers as expected the one of the $C_{w}$ at $X=L/2$ and at $X=0$'s shown in the blue and black curves of the right panel of Fig.~\ref{Fig:3}. 

The results illustrated in Fig.~\ref{Fig:3} are the most exciting prediction of our numerics: while traditional semiclassical models of back-reaction in gravitation and cosmology include the effect of the quantum emission within a mean-field theory via its average contribution to the energy-stress tensor to be included in the Einstein equations~\cite{BeiLok-CondSpacetime,Robertson-PRD-PreHeatAn-2019,Pla:PRD2021}, here we see how the large fluctuations of the two-mode-squeezed-like state of the emitted $g,d$ fields directly transfer into an analogous fluctuation of the back-reaction-induced friction force. While the possibility of such an effect was implicitly mentioned in~\cite{Kofman:PRD97} and evidence of it was already visible in our single-mode calculations~\cite{Butera-BR_DCE-2019-EPL}, here we show how the consequences of the friction-induced decoherence can be dramatic in a spatially-extended multi-mode geometry.

\vspace{2mm}
\emph{Conclusions. }
In this work, we have theoretically considered an atomic Bose-Einstein condensate as an analog model of the nonlinear non-equilibrium dynamics of the inflaton field in the  pre-heating stage of the early Universe. Our numerical results predict a crucial role of quantum fluctuations in the back-reaction effect of particle production onto the inflaton field: not only does the emission of dipole and Golstone excitations leads to an effective damping of the breathing mode as predicted by a semiclassical picture, but is also responsible for a quick decoherence of its initially in-phase excitation. The generality of the microscopic processes underlying our numerically observed results highlights the importance of going beyond semiclassical approaches \cite{Sanders-prd-2015,Pla:PRD2021} and including quantum fluctuation features in the description of back-reaction phenomena in gravitation and cosmology. Future work will extend the study to the elusive back-reaction phenomena that are responsible for black hole evaporation under the effect of Hawking emission.

\vspace{2mm}
\emph{Acknowledgements.} Continuous support from Massimiliano Rinaldi on cosmological issues is most appreciated, as well as enlightening discussions with Bei-Lok Hu and Renaud Parentani. S.~B. acknowledges funding from the Leverhulme Trust Grant No. ECF-2019-461, and from University of Glasgow via the Lord Kelvin/Adam Smith (LKAS) Leadership Fellowship. I.C. acknowledges support from the European Union Horizon 2020 research and innovation program under Grant Agreement No. 820392 (PhoQuS) and from the Provincia Autonoma di Trento.

\bibliography{QFTCS-BR.bib}
\end{document}